\documentclass[12pt]{article}
\usepackage[centertags]{amsmath}
\usepackage{amsfonts}
\usepackage{amssymb}
\usepackage{amsthm}
\newcommand{\beq}{\begin{equation}}
\newcommand{\eeq}[1]{\label{#1}\end{equation}}
\newcommand{\bea}{\begin{eqnarray}}
\newcommand{\eea}[1]{\label{#1}\end{eqnarray}}

\newcommand{\Tr}{{\rm Tr}\,}
\begin{document}
\setlength{\topmargin}{-1cm} \setlength{\oddsidemargin}{0cm}
\setlength{\evensidemargin}{0cm}
\begin{titlepage}
\begin{center}
{\Large \bf Helicity-1/2 Mode as a Probe of\\Interactions of Massive Rarita-Schwinger Field}

\vspace{20pt}

{\large Rakibur Rahman}

\vspace{12pt}

Physique Th\'eorique et Math\'ematique \& International Solvay Institutes\\
Universit\'e Libre de Bruxelles, Campus Plaine C.P. 231, B-1050 Bruxelles, Belgium

\end{center}
\vspace{20pt}

\begin{abstract}

We consider the electromagnetic and gravitational interactions of a massive Rarita-Schwinger field. St\"uckelberg analysis
of the system, when coupled to electromagnetism in flat space or to gravity, reveals in either case that the effective
field theory has a model-independent upper bound on its UV cutoff, which is finite but parametrically larger than the
particle's mass. It is the helicity-1/2 mode that becomes strongly coupled at the cutoff scale. If the interactions are
inconsistent, the same mode becomes a telltale sign of pathologies. Alternatively, consistent interactions are those
that propagate this mode within the light cone. Studying its dynamics not only sheds light on the Velo-Zwanziger
acausality, but also elucidates why supergravity and other known consistent models are pathology-free.

\end{abstract}

\end{titlepage}

\newpage

\section{Introduction}

The Rarita-Schwinger field carries a spin-3/2 representation of the Poincar\'e group, whose non-interacting
massive theory is described by the following Lagrangian~\cite{RS}: \beq \mathcal{L}_{\text{free}}=-i\bar
{\psi}_\mu\gamma^{\mu\nu\rho}\partial _\nu\psi_\rho - im\bar{\psi}_\mu\gamma^{\mu\nu}\psi_\nu,\eeq{i1}
$m$ being the mass\,\footnote{Our conventions are that the metric is mostly positive, the Clifford algebra
is $\{\gamma^\mu,\gamma^\nu\}=+2g^{\mu\nu}$, $\gamma^{\mu\,\dagger}=\eta^{\mu\mu}\gamma^\mu$, $\gamma_5
=-i\gamma^0\gamma^1\gamma^2\gamma^3$, $\gamma^{\mu_1....\mu_n}= \tfrac{1}{n!}\gamma^{\mu_1}\gamma^{\mu_2}...
\gamma^{\mu_n}+\mbox{antisymmetrization}$. The Dirac adjoint is defined as $\bar{\psi}_\mu=\psi^\dagger_\mu
\gamma^0$. The totally antisymmetric tensor $\epsilon_{\mu\nu\rho\sigma}$ is normalized as $\epsilon_{0123}=+1$.}.
The Dirac equation: $(\displaystyle{\not{\!\partial\!\,}} - m)\psi_\mu=0$, along with the correct constraints:
$\partial^\mu\psi_\mu=\gamma^\mu\psi_\mu=0$, can easily be reproduced from the Lagrangian equations of motion.
The degrees of freedom count works as follows. In 4D the vector-spinor $\psi_\mu$ contains $4\times4=16$ components.
The transversality and $\gamma$-tracelessness constraints each remove 4 of them, so that one is left with 8 degrees
of freedom (4 field variables plus 4 conjugate momenta). Indeed, a massive  spin-3/2 particle has 4
physical polarizations.

When interactions are turned on, as noticed by various authors~\cite{VZ,Misc,DW}, the theory is generically
fraught with inconsistencies even at the classical level\,\footnote{Pathologies at the quantum level were noticed
much earlier in~\cite{JS}, where it was shown that canonical commutators may become ill-defined in an interacting
theory.}, despite the fact that one starts from a Lagrangian as per suggestions made in~\cite{FP}. The interacting
theory may fail to reproduce the necessary constraints that forbid propagating unphysical modes or may give rise
to the Velo-Zwanziger acausality~\cite{VZ}, i.e. allow faster-than-light speeds for the physical modes.
The addition of non-minimal terms and/or new dynamical fields may come to the rescue. For example, the Lagrangian
proposed in~\cite{PR2} incorporates appropriate non-minimal terms that only causally propagate the physical modes
of a massive spin-3/2 in a constant external electromagnetic (EM) background. A more well-known example is
$\mathcal{N}=2$ (broken) supergravity~\cite{SUGRA1,SUGRA2}, which contains a massive gravitino that propagates
consistently, even when the cosmological constant is set to zero, given that it has a charge,
$e=\tfrac{1}{\sqrt{2}}(m/M_\text{P})$, under the graviphoton~\cite{DZ}. Here causality is preserved by the
presence of \emph{both} EM and gravity, along with non-minimal terms.

The pathologies arising in an interacting theory are due to a simple fact: the kinetic part of the free theory~(\ref{i1})
enjoys a gauge invariance, and the zero modes may acquire non-vanishing but non-canonical kinetic terms in the presence
of interactions. The best way of understanding these issues is the St\"uckelberg formalism, which was employed in the
context of massive spin 2, for example, in~\cite{Nima,PR0}. To understand this formalism, let us notice that in the
\emph{massive} theory~(\ref{i1}) gauge invariance can be restored by introducing a spin-1/2 (St\"uckelberg) field $\chi$
through the field redefinition: \beq \psi_\mu\rightarrow\psi'_\mu=\psi_\mu-\frac{1}{m}\partial_\mu\chi.\eeq{i2}
Now the Lagrangian is manifestly invariant under the St\"uckelberg symmetry: \beq \delta\psi_\mu=\partial_\mu\lambda,
\qquad \delta\chi=m\lambda,\eeq{i3} where $\lambda$ is a fermionic gauge parameter. Note that when the field redefinition~(\ref{i2})
is implemented, potentially bad higher-derivative terms in $\chi$ are killed by the antisymmetry of $\gamma^{\mu\nu}$. This
is a slick way of understanding the structure of the mass term in~(\ref{i1}).

The St\"uckelberg field is a mere redundancy since one can always choose a gauge in which $\chi=0$, as in the Lagrangian~(\ref{i1}).
The unitary gauge, however, obscures the subtleties associated with an interacting theory, and is therefore not particularly
illuminating when interactions are present. On the other hand, as we will see, the intricacies become rather transparent in
a different, judiciously chosen, gauge that instead renders the kinetic operators diagonal. For an interacting theory,
the latter gauge choice enables one to assign canonical dimensions to potential non-renormalizable operators.

The organization of this paper is as follows. In the Section 2 we consider minimal EM and gravitational
couplings of a massive Rarita-Schwinger field, and show that each theory possesses an intrinsic finite UV cutoff,
which can be improved neither by field redefinitions nor by the addition of non-minimal terms. In Section~3 we perform
St\"uckelberg analysis of various (in)consistent Lagrangians that attempt to describe interactions of a massive
spin-3/2 field. In particular, Section 3.1 considers minimal EM coupling and reproduces the Velo-Zwanziger
result~\cite{VZ}, while Section 3.2 sheds a new light on why the non-minimal Lagrangian presented in~\cite{PR2}
is consistent. Section 3.3 reconfirms that minimal gravitational coupling is pathology-free in arbitrary Einstein
spaces~\cite{Buch}, and finally Section 3.4 analyzes the consistency of $\mathcal{N}=2$ (broken)
supergravity~\cite{SUGRA1,SUGRA2,DZ}. We conclude with some remarks in Section 4.

\section{Ultraviolet Cutoff}

Local Lagrangians describing the interactions of a massive spin-3/2 field do not have a smooth massless limit.
Because the free part of the Lagrangian acquires a gauge invariance in this limit, propagators of
the massive theory become singular, so that scattering amplitudes diverge. Notice, however, that if
we introduce minimal coupling (to EM or gravity) in the Rarita-Schwinger action~(\ref{i1}), no inverse
powers of the mass appear in the resulting Lagrangian. Thus the massless singularity is not at all obvious
in the unitary gauge.

The St\"uckelberg formalism, on the other hand, focuses precisely on the gauge modes responsible for bad high
energy behavior. One can ``invent'' the St\"uckelberg symmetry and then exploit it to make a judicious covariant
gauge fixing such that the propagators acquire smooth massless limit. In this gauge one will end up having
explicit dependence on inverse powers of the mass in the form of non-renormalizable interaction terms that
involve the St\"uckelberg field $\chi$. The cutoff scale can be read off from the most divergent terms in
the Lagrangian $-$ the terms that survive in an appropriate scaling limit of zero mass and zero coupling.

\subsection{EM Coupling in Flat Space}

First we consider EM coupling in flat space, and show that the theory has an upper bound on its UV
cutoff\,\footnote{This was originally considered in~\cite{PR1}. Here we reconsider it, with a more refined analysis,
for the sake of completeness. The analysis will also be useful for the latter parts of the paper.}. When minimally
coupled to a $U(1)$ gauge field, the St\"uckelberg invariant Lagrangian for a massive Rarita-Schwinger
field reads \beq \mathcal L_{\text{em}}=-i\left(\bar{\psi}_\mu-\frac{1}{m}\bar{\chi}\overleftarrow{D}_\mu\right)
\left(\gamma^{\mu\nu\rho}D_\nu+m\gamma^{\mu\rho}\right)\left(\psi_\rho-\frac{1}{m}D_\rho\chi\right)-\tfrac{1}{4}
F_{\mu\nu}^2,\eeq{em1} which has the manifest gauged St\"uckelberg symmetry: \beq \delta\psi_\mu=D_{\mu}\lambda,
\qquad \delta\chi=m\lambda,\eeq{em2} where the covariant derivatives obey $[D_\mu, D_\nu]=ieF_{\mu\nu}$.
More explicitly, \beq \mathcal L_{\text{em}}=\mathcal L_{3/2}+\mathcal L_{\text{mix}}+\mathcal L_{\text{int}}
-\tfrac{1}{4}F_{\mu\nu}^2,\eeq{em3} where $\mathcal L_{3/2}$ involves only the helicity-3/2 mode, $\mathcal L_{\text{mix}}$
is the kinetic mixing between the two modes, and $\mathcal L_{\text{int}}$ are non-renormalizable
interaction terms, respectively given as \bea \mathcal L_{3/2}&=&-i\bar{\psi}_\mu\gamma^{\mu\nu\rho}D_\nu\psi_\rho
-im\bar{\psi}_\mu\gamma^{\mu\nu}\psi_\nu,\label{em4}\\\mathcal L_{\text{mix}}&=&i(\bar{\psi}_\mu\gamma^{\mu\nu}
D_\nu\chi+\bar{\chi}\overleftarrow{D}_\mu\gamma^{\mu\nu}\psi_\nu),\label{em5}\\\mathcal L_{\text{int}}&=&
\frac{e}{2m}\,F_{\mu\nu}(\bar{\chi}\gamma^{\mu\nu\rho}\psi_\rho-\bar{\psi}_\rho\gamma^{\mu\nu\rho}\chi
-\bar{\chi}\gamma^{\mu\nu}\chi)-\frac{e}{2m^2}\,F_{\mu\nu}\bar{\chi}\gamma^{\mu\nu\rho}D_\rho\chi.\eea{em6}
The kinetic mixing can be removed by a field redefinition, namely \beq \psi_\mu\rightarrow\psi_\mu+\tfrac{1}{2}
\gamma_\mu\chi,\eeq{em7} which, at the same time, produces a kinetic term for $\chi$ as well as mass mixing.
Now we can add the following gauge-fixing term to the Lagrangian: \beq \mathcal{L}_{\text{gf}}=i\bar{\psi}_\mu
(\gamma^{\mu\nu}\gamma^\rho-\gamma^{\mu}\eta^{\nu\rho})D_\nu\psi_\rho+im\bar{\psi}_\mu\gamma^\mu\gamma^\nu
\psi_\nu+\tfrac{3}{2}im(\bar{\psi}_\mu\gamma^\mu\chi-\bar{\chi}\gamma^\mu\psi_\mu-\bar{\chi}\chi),\eeq{em8}
which renders the propagators smooth in the massless limit, thanks to the identity \beq \gamma^{\mu\nu\rho}
=\gamma^{\mu\nu}\gamma^\rho+\eta^{\mu\rho}\gamma^\nu-\eta^{\nu\rho}\gamma^\mu.\eeq{id1} The same removes the
mass mixing as well, finally giving \bea \mathcal{L}_{\text{em}}&=&-i\bar{\psi}_\mu(\not{\!\!D}-m)\psi^\mu
-\tfrac{3}{2}\,i\bar{\chi}(\not{\!\!D}-m)\chi-\tfrac{1}{4}F_{\mu\nu}^2\nonumber\\&&+\frac{e}{2m}\,F_{\mu\nu}
(\bar{\chi}\gamma^{\mu\nu\rho}\psi_\rho-\bar{\psi}_\rho\gamma^{\mu\nu\rho}\chi+\bar{\chi}\gamma^{\mu\nu}\chi)
-\frac{e}{2m^2}\,F_{\mu\nu}(\bar{\chi}\gamma^{\mu\nu\rho}D_\rho\chi).\eea{em9}

For $e\ll1$, the most dangerous terms in the high energy limit are the dimension-6 operators. Note that the
degree of divergence does not improve with the addition of non-minimal terms, since any such operator is necessarily
irrelevant. Even a dipole term, \beq \mathcal L_{\text{dipole}}=\frac{ea}{m}\,F^{\mu\nu}\bar{\psi}_\mu
\psi_\nu\rightarrow\frac{ea}{m}\,F^{\mu\nu}\left(\bar{\psi}_\mu-\frac{1}{2}\bar{\chi}\gamma_\mu-\frac{1}{m}
\bar{\chi}\overleftarrow{D}_\mu\right)\left(\psi_\nu+\frac{1}{2}\gamma_\nu\chi-\frac{1}{m}D_\nu\chi\right),\eeq{em10}
introduces, among others, equally bad but new dimension-6 operators that involve both the helicities. Clearly,
higher multipole operators will worsen the degree of divergence\,\footnote{We emphasize that here we are only 
attempting to improve the degree of divergence, as we are looking for a theoretical upper bound on the cutoff scale that no
theory can beat. In no way do we mean that non-minimal terms are forbidden. In fact, they do appear in consistent models,
e.g., supergravity. But then the theory will have a cutoff that is simply lower than the upper bound we are trying
to find.}. Now one can take the scaling limit: $m\rightarrow0$ and $e\rightarrow0$, such that
$m^2/e\equiv\Lambda^2_{\text{em}}$=constant. The Lagrangian then reduces, after the rescaling
$\chi\rightarrow\sqrt{\tfrac{2}{3}}\,\chi$, to \beq \mathcal L_{\text{em}}\rightarrow
-i\bar{\psi}_\mu\not{\!\partial\!\,}\psi^\mu-i\bar{\chi}\not{\!\partial\!\,}\chi-\tfrac{1}{4}F_{\mu\nu}^2
-\frac{1}{3\Lambda^{2}_{\text{em}}}\,F_{\mu\nu}(\bar{\chi}\gamma^{\mu\nu\rho}\partial_\rho\chi).\eeq{em11}

Notice, however, that the non-renormalizable operators in~(\ref{em11}) are all proportional to the
equations of motion, up to total derivatives. Indeed, one can use the identity~(\ref{id1}) to write
\beq F_{\mu\nu}(\bar{\chi}\gamma^{\mu\nu\rho}\partial_\rho\chi)=\tfrac{1}{2}F_{\mu\nu}(\bar{\chi}\gamma^{\mu\nu}
\not{\!\partial\!\,}\chi-\bar{\chi}\overleftarrow{\not{\!\partial\!\,}}\gamma^{\mu\nu}\chi)-\partial_\mu F^{\mu\nu}
(\bar{\chi}\gamma_\nu\chi).\eeq{em12} Therefore, one can eliminate them by appropriate field redefinitions of $\chi$
and $A_\mu$, namely \beq \chi\rightarrow\chi+\frac{i}{6\Lambda^{2}_{\text{em}}}\,F_{\mu\nu}\gamma^{\mu\nu}\chi,\qquad
A_\mu\rightarrow A_\mu-\frac{1}{3\Lambda^{2}_{\text{em}}}\,\bar{\chi}\gamma_\mu\chi,\eeq{em13} as canceling
contributions come from the helicity-1/2 and photon kinetic terms. The price one has to pay is that new non-renormalizable
operators of dimensions 8, 10 and 12 show up, all with various negative powers of the scale $\Lambda_{\text{em}}$.
Can we add local counter-terms to the original action, which eliminate all these operators up to total derivatives,
and introduce only new terms that vanish in the above scaling limit? A positive answer would mean that one may
improve the degree of divergence of the minimally coupled theory by field redefinitions plus the addition of local
counter-terms. To see that this is not the case, let us consider the dimension-8 operator $(\bar{\chi}\gamma^\mu\chi)
\Box(\bar{\chi}\gamma_\mu\chi)$, which comes from the photon-field redefinition acting on the last term of Eq.~(\ref{em12}).
It is neither proportional to the equations of motion nor does it contain the EM field strength. Without worsening the
degree of divergence, such operators may only be produced by 4-Fermi-like local counter-terms, which in the unitary gauge
look like $(e^2/m^2)\bar{\psi}\psi\bar{\psi}\psi$. More explicitly, \bea \mathcal L_{\text{c.t.}}&\rightarrow&
b\left(\frac{e}{m}\right)^2\left(\bar{\psi}_\mu-\sqrt{\tfrac{1}{6}}\,\bar{\chi}\gamma_\mu-\frac{1}{m}\sqrt{\tfrac{2}{3}}\,\bar{\chi}
\overleftarrow{D}_\mu\right)\gamma^{\mu\nu\rho\sigma}\left(\psi_\nu+\sqrt{\tfrac{1}{6}}\,\gamma_\nu\chi-\frac{1}{m}\sqrt{\tfrac{2}{3}}\,
D_\nu\chi\right)\nonumber\\&&\times\left(\bar{\psi}_\rho-\sqrt{\tfrac{1}{6}}\,\bar{\chi}\gamma_\rho-\frac{1}{m}
\sqrt{\tfrac{2}{3}}\,\bar{\chi}\overleftarrow{D}_\rho\right)\left(\psi_\sigma+\sqrt{\tfrac{1}{6}}\,\gamma_\sigma\chi-\frac{1}{m}
\sqrt{\tfrac{2}{3}}\,D_\sigma\chi\right)+~...~,\eea{em15} where $\gamma^{\mu\nu\rho\sigma}$ plays the essential role of killing
the more dangerous operators. However, such counter-terms produce $-$ on top of those that we want to eliminate $-$ new dimension-8 
operators involving \emph{both} helicities that survive in the scaling limit.

Thus the effective field theory of a massive Rarita-Schwinger field interacting with EM in flat space has a finite
intrinsic upper bound on its cutoff: \beq \Lambda_{\text{em}}=\frac{m}{\sqrt{e}}\,,\eeq{em16} which is parametrically
larger than $m$. As seen from Eq.~(\ref{em11}), the breakdown of the effective action is due to the helicity-1/2 mode $\chi$
that becomes strongly coupled at high energies.

\subsection{Gravitational Coupling}

The St\"uckelberg invariant action for a massive spin-3/2 field minimally coupled to gravity is
\beq \mathcal L_{\text{g}}=-i\sqrt{-g}\left(\bar{\psi}_\mu-\frac{1}{m}\bar{\chi}\overleftarrow{\nabla}_\mu\right)
\left(\gamma^{\mu\nu\rho}\nabla_\nu+m\gamma^{\mu\rho}\right)\left(\psi_\rho-\frac{1}{m}\nabla_\rho\chi\right)
+\tfrac{1}{2}M^2_\text{P}\sqrt{-g}R.\eeq{g1} Here the commutator of the covariant derivatives acts on different modes as:
\bea [\nabla_\mu, \nabla_\nu]\psi_\rho&=&-R_{\mu\nu\rho}^{~~~~\sigma}\psi_\sigma+\tfrac{1}{4}R_{\mu\nu\alpha\beta}
\gamma^{\alpha\beta}\psi_\rho,\label{g2}\\~[\nabla_\mu, \nabla_\nu]\chi&=&\tfrac{1}{4}R_{\mu\nu\alpha\beta}
\gamma^{\alpha\beta}\chi.\eea{g3} One can work out the Lagrangian~(\ref{g1}) to write
\beq \mathcal L_{\text{g}}=\mathcal L_{3/2}+\mathcal L_{\text{mix}}+\mathcal L_{\text{int}}
+\tfrac{1}{2}M^2_\text{P}\sqrt{-g}R,\eeq{g4} where $\mathcal L_{3/2}$ and $\mathcal L_{\text{mix}}$ are the
gravitational counterparts of those given by Eqs.~(\ref{em4}) and~(\ref{em5}) respectively, while
$\mathcal L_{\text{int}}$ are the non-renormalizable interactions. The latter can be computed explicitly,
using Eqs.~(\ref{g2})-(\ref{g3}), the Bianchi identity: $R_{[\mu\nu\alpha]\beta}=0$, and various $\gamma$-matrix
identities. The following ones are particularly useful: \beq \gamma^{\mu\nu\rho}\gamma^{\alpha\beta}
R_{\mu\nu\alpha\beta}(\psi_\rho,\nabla_\rho\chi)=4G^{\mu\nu}\gamma_\mu(\psi_\nu,\nabla_\nu\chi),\qquad
\gamma^{\mu\nu}\gamma^{\alpha\beta}R_{\mu\nu\alpha\beta}=-2R,\eeq{g5} where $G^{\mu\nu}$ is the Einstein tensor.
The result is \beq \mathcal L_{\text{int}}=-\frac{i}{2m}\,\sqrt{-g}\left[\,G^{\mu\nu}\left(\bar{\chi}\gamma_\mu
\psi_\nu-\bar{\psi}_\mu\gamma_\nu\chi\right)+\tfrac{1}{2}\bar{\chi}R\chi\,\right]+\frac{i}{2m^2}\,\sqrt{-g}\,
G^{\mu\nu}\bar{\chi}\gamma_\mu\nabla_\nu\chi.\eeq{g6}

The field redefinition that eliminates the kinetic mixing is the same as Eq.~(\ref{em7}), while the desired
gauge-fixing term is just the gravitational counterpart of Eq.~(\ref{em8}). One is left with
\bea \mathcal{L}_{\text{g}}&=&-i\sqrt{-g}\left[\,\bar{\psi}_\mu(\not{\!\nabla\!}-m)\psi^\mu
+\tfrac{3}{2}\,\bar{\chi}(\not{\!\nabla\!}-m)\chi\,\right]+\tfrac{1}{2}M^2_\text{P}\sqrt{-g}R\nonumber\\&&
-\frac{i}{2m}\,\sqrt{-g}\left[\,G^{\mu\nu}\left(\bar{\chi}\gamma_\mu\psi_\nu-\bar{\psi}_\mu\gamma_\nu\chi\right)
-\tfrac{1}{2}\bar{\chi}R\chi-\frac{1}{m}\,G^{\mu\nu}\bar{\chi}\gamma_\mu\nabla_\nu\chi\,\right].\eea{g7}
Before assigning canonical dimensions to various operators, we must canonically normalize the graviton field
$h_{\mu\nu}$, so that it has mass dimension one: \beq g_{\mu\nu}=\eta_{\mu\nu}+\frac{1}{M_\text{P}}
h_{\mu\nu}.\eeq{g8} We take $m\ll M_\text{P}$, which is essential for a sensible effective field theory to exist.
We see that in the high energy limit the most dangerous terms are the dimension-7 operators contained in
$G^{\mu\nu}\bar{\chi}\gamma_\mu\nabla_\nu\chi$, which are $\chi-h-\chi$ vertices. Because non-minimal interactions
show up with Planck-mass suppression in the unitary gauge, they can contribute only less divergent terms to the
Lagrangian~(\ref{g7}). Thus they are harmless, but they do not improve the degree of divergence either.

The high energy regime we are interested in $-$ characterized by the center-of-mass energy 
$m\ll\sqrt s\ll M_\text P$ $-$ includes two parametrically disparate scales of interest,
\beq \Lambda_*\equiv\sqrt[3]{m^2M_{\text P}}\,,\qquad \Lambda_{\text{g}}\equiv\sqrt{mM_{\text P}}\,,\eeq{scales}
where $\Lambda_*\ll\Lambda_{\text{g}}$. Now, with the rescaling $\chi\rightarrow\sqrt{\tfrac{2}{3}}\,\chi$,
our Lagrangian~(\ref{g7}) reduces to \beq \mathcal L_{\text{g}}\rightarrow-i\bar{\psi}_\mu\left(\not{\!\partial\!\,}
-m\right)\psi^\mu-i\bar{\chi}\left(\not{\!\partial\!\,}-m\right)\chi+h_{\mu\nu}\mathcal{G}^
{\mu\nu}+\frac{i}{3\Lambda^3_*}\,\mathcal G^{\mu\nu}\bar{\chi}\gamma_\mu\partial_\nu\chi+...\,,\eeq{g9}
where the ellipses stand for less divergent terms that become important at scales $\Lambda_{\text{g}}$ or higher.
Here $\mathcal{G}^{\mu\nu}\equiv(\mathcal E\cdot h)^{\mu\nu}$ is the linearized Einstein tensor, and
\beq \mathcal E^{\mu\nu\alpha\beta}=\tfrac{1}{2}\left[\,\left(\eta^{\mu\nu,\alpha\beta}
-\eta^{\mu\nu}\eta^{\alpha\beta}\right)\Box+\eta^{\mu\nu}\partial^\alpha\partial^\beta+\eta^{\alpha\beta}
\partial^\mu\partial^\nu-\eta^{\mu(\alpha}\partial^{\beta)}\partial^\nu-\eta^{\nu(\alpha}\partial^{\beta)}
\partial^\mu\,\right],\eeq{g10} so that
$h_{\mu\nu}\mathcal{G}^{\mu\nu}$ is the kinetic term for the canonically normalized graviton $h_{\mu\nu}$.
It is clear that the dimension-7 operator in Eq.~(\ref{g9}) can be eliminated by the field redefinition:
\beq h_{\mu\nu}\rightarrow h_{\mu\nu}-\frac{i}{6\Lambda^3_*}\,\bar{\chi}\gamma_{(\mu}\partial
_{\nu)}\chi.\eeq{g11} But this will leave us with the following dimension-10 operator, quartic in $\chi$:
\beq \mathcal L_{\text{dim-10}}=-\frac{1}{36\Lambda^6_*}\,(\bar{\chi}\gamma_\mu\partial_\nu\chi)
\mathcal E^{\mu\nu\alpha\beta}(\bar{\chi}\gamma_\alpha\partial_\beta\chi).\eeq{g12} This is a contact
term for 4 helicity-1/2 modes. Because we are interested in on-shell scattering amplitudes, some pieces
contained in Eq.~(\ref{g12}) may actually be less divergent, thanks to the equation of motion
$\displaystyle\not{\!\!\partial\!\,}\chi=m\chi+...$\,. Indeed, all but the first term from the expression~(\ref{g10})
for $\mathcal E^{\mu\nu\alpha\beta}$ give $-$ up to total derivatives $-$ dimension-8 operators that go like
$1/\Lambda_\text{g}^4$. This follows partly from the fact that, unlike in the electromagnetic case, here one is
dealing with Majorana fermions, so that one has $\bar\chi\gamma^\mu\chi=0$. Thus one is left with
\beq \mathcal L_{\text{int}}\rightarrow-\frac{1}{72\Lambda^6_*}\,(\bar{\chi}\gamma_\mu\partial_\nu\chi)\,
\eta^{\mu\nu,\alpha\beta}\,\Box(\bar{\chi}\gamma_\alpha\partial_\beta\chi)+...\,.\eeq{g13}
This operator does not reduce further for on-shell $\chi$. However, as we will see, it can be canceled, up to
total derivatives, by addition of local counter-terms.

In the unitary gauge, the potentially interesting counter-terms are 4-Fermi interactions:
\beq \mathcal L_{\text{c.t.}}=M_{\text P}^{-2}(\bar{\psi}_\rho\gamma_\mu\psi_\nu)\mathcal A^{\mu\nu\alpha\beta\rho\sigma}
(\bar{\psi}_\sigma\gamma_\alpha\psi_\beta),\eeq{ct1} where $\mathcal A^{\mu\nu\alpha\beta\rho\sigma}$ is a
dimensionless tensor. The replacement $\psi_\mu\rightarrow\psi_\mu+\sqrt{\tfrac{1}{6}}\,\gamma_\mu\chi-
\sqrt{\tfrac{2}{3}}\,\partial_\mu\chi/m$ will then give rise to dimension-10 operators, quartic in $\chi$,
which may cancel those of Eq.~(\ref{g13}). It is easy to find that the required cancelation takes place for
\beq \mathcal A^{\mu\nu\alpha\beta\rho\sigma}=\frac{1}{32}\left(\eta^{\mu\alpha}\eta^{\nu[\beta}
\eta^{\sigma]\rho}+2\eta^{\alpha[\nu}\eta^{\rho][\sigma}\eta^{\beta]\mu}\right).\eeq{ct2}
Note that the antisymmetry in the indices $(\rho,\nu)$ and $(\sigma,\beta)$ ensures that no new dimension-10
operators are generated. Thus no terms remain that become important at $\Lambda_*$\,:
the counter-term~(\ref{ct1}) has improved the high energy behavior of the system.

Next, one would like to consider the dimension-9 operators coming from this counter-term that blow up at the
scale $\sqrt{\Lambda_*\Lambda_\text g}$ $-$ higher than $\Lambda_*$ but lower than $\Lambda_\text g$.
A straightforward computation shows that all such contact terms are actually less divergent for on-shell $\chi$.
Therefore, the strong coupling regime is pushed even higher to the scale $\Lambda_\text g$.

Can we improve the cutoff scale any further? The answer is negative. To see this, let us take the scaling limit:
$m\rightarrow0$ and $M_{\text P}\rightarrow\infty$, such that $\Lambda_{\text{g}}=$ constant. This gives
\beq \mathcal L_{\text{g}}+\mathcal L_{\text{c.t.}}\rightarrow-i\bar{\psi}_\mu\not{\!\partial\!\,}\psi^\mu
-i\bar{\chi}\not{\!\partial\!\,}\chi+h_{\mu\nu}\mathcal{G}^{\mu\nu}-\frac{i}{\Lambda^2_\text{g}}\,\mathcal G^{\mu\nu}
\left(\bar{\chi}\gamma_\mu\psi_\nu+\tfrac{1}{4}\eta_{\mu\nu}\bar{\chi}\chi\right)+\mathcal L_\text{dim-8}\,,\eeq{final}
where the dimension-8 operators, which are $\mathcal O(1/\Lambda^4_\text g)$, contain quartic contact terms originating
from the naive dimension-10 operator~(\ref{g12}) as well as from the counter-term~(\ref{ct1}). Another field redefinition
of the graviton, namely \beq h_{\mu\nu}\rightarrow h_{\mu\nu}+\frac{i}{2\Lambda^2_\text{g}}\left(\bar{\chi}\gamma_{(\mu}
\psi_{\nu)}+\tfrac{1}{4}\eta_{\mu\nu}\bar{\chi}\chi\right)\eeq{final2} will remove all the dimension-6 operators
in~(\ref{final}) and give rise to additional dimension-8 operators. For simplicity let us look at all the dimension-8
quartic terms that involve 2 helicity-3/2 and 2 helicity-1/2 modes. They are
\bea \Lambda^4_\text{g}\,\mathcal L_{\psi\psi\chi\chi}&=&\tfrac{2}{3}\,(\bar{\psi}_\rho\gamma_\mu\psi_\nu)
\mathcal A^{\mu\nu\alpha\beta\rho\sigma}(\partial_\sigma\bar{\chi}\gamma_\alpha\partial_\beta\chi)+\tfrac{2}{3}\,
(\partial_\rho\bar{\chi}\gamma_\mu\partial_\nu\chi)\mathcal A^{\mu\nu\alpha\beta\rho\sigma}(\bar{\psi}_\sigma
\gamma_\alpha\psi_\beta)\nonumber\\&&+\tfrac{8}{3}\,(\bar{\psi}_\rho\gamma_\mu\partial_\nu\chi)
\mathcal A^{\mu\nu\alpha\beta\rho\sigma}(\bar{\psi}_\sigma\gamma_\alpha\partial_\beta\chi) -\tfrac{1}{4}\,
\left(\bar{\chi}\gamma_\mu\psi_\nu\right)\mathcal E^{\mu\nu\alpha\beta}\left(\bar{\chi}\gamma_\alpha\psi_\beta\right).\eea{final3}
Notice that the last term that comes from the field redefinition~(\ref{final2}) contains pieces that are non-vanishing on-shell.
Can these be canceled by the first three terms? No, because of simple symmetry considerations. The latter set of terms enjoys
the shift of $\chi$ by a constant spinor, while the former does not. At this point, we also have exhausted the possibility
of local counter terms coming to the rescue.

Thus we have found an upper bound on the UV cutoff of the effective theory describing a gravitationally interacting
massive spin-3/2 field: \beq \Lambda_{\text{g}}=\sqrt{mM_{\text P}}\,.\eeq{g100} This is finite but parametrically
larger than the mass. Again, it is the helicity-1/2 mode that is responsible for the strong coupling around the cutoff scale.

Our result\,\footnote{This result agrees with the conjecture $\Lambda_\text g=\left(m^{2s-2}M_\text{P}\right)^{1/(2s-1)}$
for generic spin $s$ made in Ref.~\cite{Thesis}.} is hardly a surprise given the existence of $\mathcal N=1$ broken
supergravity~\cite{Cremmer:1978hn}. This theory possesses remarkably good properties in the high energy limit, and its
strong coupling regime has been investigated in Ref.~\cite{sugrastrong}. When the (pseudo)scalars are decoupled from the theory,
with their masses sent to infinity, one ends up having only a massive gravitino coupled to gravity. This is the system we have considered
in this Section, and indeed the theory has a cutoff around the supersymmetry breaking scale $\Lambda_\text g$~\cite{sugrastrong}.

\section{Interacting Theories of a Rarita-Schwinger Field}

Now we will consider various (in)consistent models of an interacting massive spin-3/2 field, and analyze them through
the St\"uckelberg formalism. As we already know, when interactions are present, the helicity-1/2 mode generally acquires
non-standard kinetic terms. In inconsistent theories this mode may move faster than light or even cease to propagate.
The consistency of interacting theories crucially relies on having a pathology-free helicity-1/2 sector.
Conversely, by ensuring that this mode does not exhibit pathological behavior, we can (re)derive conditions that
render a theory consistent.

\subsection{Minimal EM Interaction}

This has already been considered in Section 2.1, and we recall from Eq.~(\ref{em9}) that the Lagrangian
can be written as \bea \mathcal{L}_{\text{em}}&=&-i\bar{\psi}_\mu(\not{\!\!D}-m)\psi^\mu-\tfrac{3}{2}\,
i\bar{\chi}(\not{\!\!D}-m)\chi-\tfrac{1}{4}F_{\mu\nu}^2\nonumber\\&&+\frac{e}{2m}\,F_{\mu\nu}
(\bar{\chi}\gamma^{\mu\nu\rho}\psi_\rho-\bar{\psi}_\rho\gamma^{\mu\nu\rho}\chi+\bar{\chi}\gamma^{\mu\nu}\chi)
-\frac{e}{2m^2}\,F_{\mu\nu}(\bar{\chi}\gamma^{\mu\nu\rho}D_\rho\chi).\eea{min1} It is manifest that the
helicity-3/2 sector enjoys a healthy kinetic term. On the other hand, the $\chi$-sector is tricky, because
in an external EM background the last term in~(\ref{min1}) will act like a kinetic operator. Let us write
down the equations of motion for $\chi$: \beq -i\not{\!\partial\!\,}\chi-\tfrac{1}{2}\alpha\,\gamma^{\mu\nu\rho}
F_{\mu\nu}\partial_\rho\chi+(\text{lower-derivative terms})=0,\qquad \alpha\equiv\tfrac{2}{3}\,e/m^2.\eeq{min2}
Now we will use the method of characteristic determinants~\cite{VZ} to investigate whether this system allows
propagation outside the light cone. The method consists of determining the normal, $n_\mu=(n_0,\vec{n})$, to the
characteristic hypersurfaces. We replace $\partial_\mu$ with $-in_\mu$ in the highest derivative terms
in Eq.~(\ref{min2}), and then equate the determinant $\Delta(n)$ of the resulting coefficient matrix to zero.
The system is hyperbolic $-$ if for any $\vec{n}$ $-$ the algebraic equation $\Delta(n)=0$ has real solutions for $n_0$;
then the ratio $n_0/|\vec{n}|$ gives the maximum wave speed. The required coefficient matrix is given by
\bea \mathcal M&=&-\gamma^\mu n_\mu+\tfrac{i}{2}\alpha\,\gamma^{\mu\nu\rho}F_{\mu\nu}n_\rho\nonumber\\
&=&-i\left(\begin{array}{ll}~~~~~~~~~~~~~~~~~\mathbf{0} &-\vec{\sigma}\cdot(\vec{n}+\alpha n_0\vec{B})-(n_0+\alpha\vec{n}
\cdot\vec{B})\\\vec{\sigma}\cdot(\vec{n}-\alpha n_0\vec{B})-(n_0-\alpha\vec{n}\cdot\vec{B})&~~~~~~~~~~~~~~~~~~~~\mathbf{0}
\end{array} \right).~~~~~~~\eea{min3} To compute its determinant let us assume, without loss of generality, that the
magnetic field $\vec{B}$ points in the $z$-direction, and that the 3-vector $\vec{n}$ lies on the $zx$-plane making
an angle $\theta$ with $\vec{B}$. Thus we obtain \beq \Delta(n)\equiv\det{\mathcal M}=[n_0^2-\vec{n}^2-\alpha^2\vec{B}^2
(n_0^2-\vec{n}^2\cos^2{\theta})]^2,\eeq{min4}
which vanishes for \beq \frac{n_0}{|\vec{n}|}=\sqrt{\frac{1-\alpha^2\vec{B}^2\cos^2\theta}{1-\alpha^2\vec{B}^2}}\,.\eeq{min10}
We see that the system ceases to be hyperbolic whenever $\alpha^2\vec{B}^2$ exceeds unity, i.e., when
\beq \vec{B}^2\geq\left(\frac{3m^2}{2e} \right)^2.\eeq{min11} In addition, even an infinitesimal magnetic
field will cause superluminal propagation for generic $\theta$. This is the so-called Velo-Zwanziger problem.
The pathologies are serious in that they can very well arise when the EM field invariants $\vec{B}^2-\vec{E}^2$
and $\vec{B}\cdot\vec{E}$ are non-vanishing but small (in the units of $m^4/e^2$), so that we are far away from the
regime of instabilities~\cite{instabilities} and the notion of long-lived propagating particles makes sense.

\subsection{Consistent Non-Minimal EM Couplings}

The Velo-Zwanziger acausality shows up even for the simplest possible interaction setup of a constant
external EM background. A wide class of non-minimal models~\cite{Misc} also exhibits the same pathological
features. Porrati and Rahman~\cite{PR2} wrote down a non-minimal Lagrangian, which consistently describes
a charged massive Rarita-Schwinger field exposed to a constant EM background in flat space. In the unitary
gauge it reads~\cite{PR2}: \beq \mathcal L_{\text{PR}}=-i\bar{\psi}_\mu\gamma^{\mu\nu\rho}D_\nu\psi_\rho
-im\bar{\psi}_\mu b^{\mu\nu}\psi_\nu,\eeq{nm1} where the antisymmetric tensor $b_{\mu\nu}$ contains
``corrections'' to $\gamma_{\mu\nu}$, of the form \beq b_{\mu\nu}=\gamma_{\mu\nu}+B_{\mu\nu}^{+}+\gamma^\rho
C_{\rho[\mu}\gamma_{\nu]},\qquad B^\pm_{\mu\nu}\equiv B_{\mu\nu}\mp i\gamma_5\tilde{B}_{\mu\nu},\eeq{nm2}
with $\tilde{B}_{\mu\nu}\equiv\tfrac{1}{2}\epsilon_{\mu\nu\rho\sigma}B^{\rho\sigma}$. The Lorentz tensor
$B_{\mu\nu}$ is antisymmetric, while the Lorentz tensor $C_{\mu\nu}$ is symmetric and traceless. They are respectively
imaginary and real, as implied by the Hermiticity condition, and they both vanish in the limit $F\rightarrow0$.
They are related as~\cite{PR2}: \beq C^{\mu\nu}=-\tfrac{1}{2}B^{+\mu\rho}B_\rho^{-\,\nu}
=-\tfrac{1}{2}B^{-\mu\rho}B_\rho^{+\,\nu}=-\left[B^{\mu\rho}B_\rho^{~\nu}-\tfrac{1}{4}\eta^{\mu\nu}
\text{Tr}(B^2)\right],\eeq{nm3} while $B_{\mu\nu}$ can be computed from the relation~\cite{PR2}:
\beq B_{\mu\nu}=i(e/m^2)F_{\mu\nu}+\tfrac{1}{4}\text{Tr}(B^2)B_{\mu\nu}-\tfrac{1}{4}\,\text{Tr}(B\tilde{B})
\tilde{B}_{\mu\nu},\eeq{nm4} as a power series in the EM field strength $F_{\mu\nu}$, which is always possible
in physically interesting situations, i.e., when the EM field invariants are small.

In what follows we perform a St\"uckelberg analysis of the Lagrangian~(\ref{nm1}) to reveal that the
relations~(\ref{nm3}) and~(\ref{nm4}) are precisely those that ensure a healthy helicity-1/2 sector.
We can render the Lagrangian St\"uckelberg invariant as usual, and work out the various terms to arrive at the
non-minimal counterpart of Eq.~(\ref{em3}): \bea \mathcal L_{\text{PR}}&=&-i\bar{\psi}_\mu\gamma^{\mu\nu\rho}
D_\nu\psi_\rho-im\bar{\psi}_\mu b^{\mu\nu}\psi_\nu+i(\bar{\psi}_\mu b^{\mu\nu}D_\nu\chi+\bar{\chi}\overleftarrow{D}
_\mu b^{\mu\nu}\psi_\nu)\nonumber\\&&+\frac{e}{2m}\,F_{\mu\nu}(\bar{\chi}\gamma^{\mu\nu\rho}\psi_\rho-\bar{\psi}
_\rho\gamma^{\mu\nu\rho}\chi-\bar{\chi}b^{\mu\nu}\chi)-\frac{e}{2m^2}\,F_{\mu\nu}\left(\bar{\chi}\gamma^{\mu\nu\rho}
D_\rho\chi\right).\eea{nm5} As we have already seen in the minimal theory, a redefinition of $\psi_\mu$ can
eliminate the kinetic mixing. To find such a field redefinition in the present case, let us take note of
the following identities that follow from elementary manipulations of $\gamma$-matrix algebra:
\bea &B^+_{\mu\nu}=-\tfrac{1}{4}\gamma^\rho\not{\!\!B\,}\gamma_{\rho\mu\nu}=-\tfrac{1}{4}\gamma_{\mu\nu\rho}
\not{\!\!B\,}\gamma^\rho,\qquad \not{\!\!B}\equiv\gamma^{\mu\nu}B_{\mu\nu},&\label{nm6}\\&\gamma^\alpha
C_{\alpha[\mu}\gamma_{\nu]}=-\tfrac{1}{2}\gamma_\alpha C^{\alpha\rho}\gamma_{\rho\mu\nu}=-\tfrac{1}{2}
\gamma_{\mu\nu\rho}C^{\rho\alpha}\gamma_\alpha.&\eea{nm7} Given this, it is not difficult to see that the
required field redefinition is \beq \psi_\mu\rightarrow\psi_\mu+\tfrac{1}{2}\left(\gamma_\mu-\tfrac{1}{2}
\not{\!\!B\,}\gamma_\mu-\gamma^\alpha C_{\alpha\mu}\right)\chi.\eeq{nm8} This, when implemented
in Eq.~(\ref{nm5}), will also produce new non-canonical kinetic operators for $\chi$, which add to the already
existing troublesome operator $F_{\mu\nu}\bar{\chi}\gamma^{\mu\nu\rho}D_\rho\chi$. One can also add the
gauge-fixing term~(\ref{em8}) to make manifest that the helicity-3/2 sector is hyperbolic and causal.
The result is the non-minimal counterpart of Eq.~(\ref{em9}), given by \bea \mathcal{L}_{\text{PR}}&=&
-i\bar{\psi}_\mu(\not{\!\!D}-m)\psi^\mu-\tfrac{3}{2}\,i\bar{\chi}(\not{\!\!D}-m)\chi+\frac{e}{2m}\,
F_{\mu\nu}(\bar{\chi}\gamma^{\mu\nu\rho}\psi_\rho-\bar{\psi}_\rho\gamma^{\mu\nu\rho}\chi+\bar{\chi}
b^{\mu\nu}\chi)\nonumber\\&&+\tfrac{1}{2}\,i\bar{\chi}\left[\,i(e/m^2)\gamma^{\mu\nu\rho}F_{\mu\nu}
+b^{\mu\rho}\left(\gamma_\mu-\tfrac{1}{2}\not{\!\!B\,}\gamma_\mu-\gamma^\alpha C_{\alpha\mu}\right)
+3\gamma^\rho\right]D_\rho\chi.\eea{nm9}

The key point is that we have at our disposal two functions of the EM field strength, $B_{\mu\nu}$ and
$C_{\mu\nu}$, which could be judiciously chosen so as to render the $\chi$-sector pathology-free. With this
end in view, we make the rescaling $\chi\rightarrow\sqrt{\tfrac{2}{3}}\,\chi$, and look at the helicity-1/2
kinetic-like operators, which we symbolically write as \beq \mathcal{L}_{\chi,\text{kin}}=-i\bar{\chi}\Gamma^\mu
D_\mu\chi.\eeq{nm11} If $\Gamma^\mu$ is proportional to $\gamma^\mu$ with a positive coefficient, the
$\chi$-sector will be ghost-free, hyperbolic and causal. The expression for $\Gamma^\mu$ can be  simplified to
\bea \Gamma^\mu&=&\gamma^\mu+\tfrac{1}{3}\left\{-i(e/m^2)\gamma^{\mu\nu\rho}F_{\nu\rho}+\gamma^{\mu\nu\rho}
B_{\nu\rho}+\gamma^\rho C_{\rho\nu}\left(B^{-\nu\mu}-\tfrac{1}{2}B^{+\nu\alpha}\gamma^\mu\gamma_\alpha\right)
\right\}\nonumber\\&&+\tfrac{1}{3}\gamma_\nu\left[2C^{\mu\nu}+B^{+\mu\rho}B_\rho^{-\,\nu}\right]
+\tfrac{1}{6}\gamma^\nu C_{\nu\rho}C^{\rho\sigma}\gamma^\mu\gamma_\sigma,\eea{nm12} thanks to the identities:
\beq \gamma^\mu B^+_{\mu\nu}=\tfrac{1}{2}\not{\!\!B\,}\gamma_\nu,\qquad B^+_{\mu\nu}\gamma^\nu =\tfrac{1}{2}
\gamma_\mu\not{\!\!B\,},\qquad \gamma_\mu\not{\!\!B\,}\gamma^\mu=0,\qquad \tfrac{1}{2}\left(\gamma^\mu
\not{\!\!B}+\not{\!\!B}\gamma^\mu\right)=\gamma^{\mu\nu\rho}B_{\nu\rho}.\eeq{nm13} In Eq.~(\ref{nm12})
if one sets the symmetric traceless tensor inside the brackets to zero $-$ which is nothing but the choice of
the relation~(\ref{nm3}) $-$ the entire second line becomes proportional to $\gamma^\mu$.
This is because of the identity \beq B^\pm_{\mu\rho}B^{\pm\, \rho\nu}=\tfrac{1}{2}[\,\Tr (B^2) \pm i
\gamma_5\Tr (B\tilde{B})\,]\,\delta_\mu^\nu,\eeq{nm15} which, along with Eq.~(\ref{nm3}), enables one to
write \beq \gamma^\nu C_{\nu\rho}C^{\rho\sigma}\gamma^\mu\gamma_\sigma=\tfrac{1}{4}\gamma^\nu(B^{+\,\alpha}
_\nu B^-_{\alpha\rho}B^{-\rho\lambda}B_\lambda^{+\,\sigma})\gamma^\mu\gamma_\sigma=-\tfrac{1}{8}
\left\{[\Tr(B^2)]^2+[\Tr(B\tilde{B})]^2\right\}\gamma^\mu.\eeq{nm16} Moreover, one can use
Eqs.~(\ref{nm3}) and~(\ref{nm15}), and the definitions of $B_{\mu\nu}^\pm$ and $\tilde{B}_{\mu\nu}$ to write
\beq \gamma^\rho C_{\rho\nu}\left(B^{-\nu\mu}-\tfrac{1}{2}B^{+\nu\alpha}\gamma^\mu\gamma_\alpha\right)
=-\tfrac{1}{4}\gamma^{\mu\nu\rho}\left[\Tr(B^2)B_{\nu\rho}-\Tr(B\tilde{B})\tilde{B}_{\nu\rho}\right].\eeq{nm16a}
Now in view of Eqs.~(\ref{nm3}),~(\ref{nm16}), and~(\ref{nm16a}), the expression for $\Gamma^\mu$ reduces to
\bea \Gamma^\mu&=&\left\{1-\tfrac{1}{48}[\Tr(B^2)]^2-\tfrac{1}{48}[\Tr(B\tilde{B})]^2\right\}\gamma^\mu
\nonumber\\&&+\tfrac{1}{3}\gamma^{\mu\nu\rho}\left\{-i(e/m^2)F_{\nu\rho}+B_{\nu\rho}-\tfrac{1}{4}\Tr(B^2)
B_{\nu\rho}+\tfrac{1}{4}\Tr(B\tilde{B})\tilde{B}_{\nu\rho}\right\}.\eea{nm17} This produces the same
kind of helicity-1/2 kinetic terms as the minimally-coupled theory. Clearly, the second line in the
above expression will give rise to pathologies unless it is set to zero. Then, the consistent propagation of
$\chi$ requires Eq.~(\ref{nm4}), and we are left with \beq \mathcal{L}_{\chi,\text{kin}}=-i\left\{1-\tfrac{1}{48}
[\Tr(B^2)]^2-\tfrac{1}{48}[\Tr(B\tilde{B})]^2\right\}\bar{\chi}\not{\!\!D}\chi.\eeq{nm18}
The factor appearing in the kinetic term manifestly depends on the relativistic field invariants in
such a way that it is always positive in the regimes of physical interest. Thus the mere requirement
of a healthy helicity-1/2 sector recovers the model~(\ref{nm1})-(\ref{nm4}).

\subsection{Minimal Coupling to Gravity}

As already considered in Section 2.2, minimal gravitational coupling shows up, interestingly, as one tries to
write down consistent models for a massive spin-3/2 field in Einstein space~\cite{Buch}. The Lagrangian found
in Ref.~\cite{Buch} (by using the BRST approach to higher-spin fields) boils down to the minimal Lagrangian
in the unitary gauge. It means that if we take the minimally coupled theory with the spin-3/2 field as a probe,
the consistent propagation of the helicity-1/2 mode would require that the Einstein tensor be proportional to
the metric.

The consistency of minimal gravitational coupling in Einstein spaces becomes rather obvious in the St\"uckelberg
language. We recall from Eq.~(\ref{g7}) that the minimally coupled theory, in $d=4$ dimensions, can be cast into
the following form: \bea \mathcal{L}_{\text{g}}&=&-i\sqrt{-g}\left[\,\bar{\psi}_\mu(\not{\!\nabla\!}-m)\psi^\mu
+\tfrac{3}{2}\,\bar{\chi}(\not{\!\nabla\!}-m)\chi\,\right]\nonumber\\&&-\frac{i}{2m}\,\sqrt{-g}
\left[\,G^{\mu\nu}\left(\bar{\chi}\gamma_\mu\psi_\nu-\bar{\psi}_\mu\gamma_\nu\chi\right)-\tfrac{1}{2}
\bar{\chi}R\chi-\frac{1}{m}\,G^{\mu\nu}\bar{\chi}\gamma_\mu\nabla_\nu\chi\,\right].\eea{buch1} With the
rescaling $\chi\rightarrow\sqrt{\tfrac{2}{d-1}}\,\chi$, the kinetic-like operators for $\chi$ become
\beq \mathcal{L}_{\chi,\text{kin}}=-i\sqrt{-g}\left[\,g^{\mu\nu}-\frac{1}{(d-1)m^2}\,G^{\mu\nu}\,\right]
\bar{\chi}\gamma_\mu\nabla_\nu\chi.\eeq{buch2} The above expression actually holds true even when $d$ is arbitrary.
It is clear that, if $G^{\mu\nu}$ is proportional to $g^{\mu\nu}$, the system reduces to a manifestly hyperbolic and
causal one. We must ensure, however, that the coefficient in front of $(\bar{\chi}\not{\!\!\nabla\!\,}\chi)$ is
always non-negative. Otherwise, as $\chi$ becomes a propagating ghost, there will be loss of unitarity. The
coefficient can be computed by noting that, for Einstein spaces one has \beq G^{\mu\nu}\equiv R^{\mu\nu}
-\tfrac{1}{2}g^{\mu\nu}R=-\left(\frac{d-2}{2d}\right)g^{\mu\nu}R,\eeq{buch3} which enables one to
rewrite Eq.~(\ref{buch2}) as \beq \mathcal{L}_{\chi,\text{kin}}=-i\sqrt{-g}\left[\,1+\frac{d-2}{2d(d-1)}
\left(\frac{R}{m^2}\right)\,\right]\bar{\chi}\not{\!\nabla\!\,}\chi.\eeq{buch4} Therefore, everywhere
in spacetime, the Ricci scalar must satisfy \beq \left(\frac{d-2}{2d}\right)R\geq-(d-1)m^2.\eeq{buch5}

Of special interest are constant curvature spaces, for which the left-hand side of Eq.~(\ref{buch5}) is nothing
but the cosmological constant $\Lambda$. The unitarity bound then reduces to \beq \Lambda\geq-(d-1)m^2.\eeq{buch6}
This is precisely the result of Ref.~\cite{partial} for a neutral massive spin-3/2 field in cosmological backgrounds.
The equality sign in Eq.~(\ref{buch6}) renders the field $\chi$ algebraic by setting its kinetic term  to zero,
and this corresponds to a genuinely massless spin-3/2 field in AdS~\cite{Buch,partial}.

\subsection{Supergravity}

It is well known that $\mathcal{N}=2$ gauged supergravity~\cite{SUGRA1} incorporates a consistently propagating
Rarita-Schwinger field (gravitino), which is coupled to a $U(1)$ field (graviphoton) as well as gravity with a
cosmological constant. When the cosmological constant is detuned from its supersymmetric value, $\Lambda=-3m^2$,
the resulting broken supergravity theory~\cite{SUGRA2,DZ} still propagates the massive gravitino causally,
for any unitarily allowed $\Lambda$, provided the usual mass-charge relation holds~\cite{DZ}, i.e., the gravitino
charge $e$ under the graviphoton is \beq e=\frac{1}{\sqrt{2}}\left(\frac{m}{M_{\text P}}\right).\eeq{s1}

We consider the gravitino as a probe in the dynamical Maxwell-Einstein background; the latter satisfies the
bosonic equations of motion of $\mathcal N=2$ (broken) supergravity: \beq \nabla_\mu F^{\mu\nu}=0,\qquad
G^{\mu\nu}+\Lambda g^{\mu\nu}=\frac{1}{M^2_{\text P}}\,T^{\mu\nu},\eeq{s2} where $T^{\mu\nu}$ is the
stress-energy tensor of the Maxwell field, given by \beq T^{\mu\nu}=-\tfrac{1}{2}F^{+\mu\rho}F_\rho^{-\,\nu}
=-\tfrac{1}{2}F^{-\mu\rho}F_\rho^{+\,\nu}=-\left[F^{\mu\rho}F_\rho^{~\nu}-\tfrac{1}{4}\eta^{\mu\nu}
\text{Tr}(F^2)\right].\eeq{s3} In this combined background of EM and gravitational fields, the probe
Rarita-Schwinger field is described in the unitary gauge by the following non-minimal Lagrangian:
\beq \mathcal L_{\text{gravitino}}=-i\sqrt{-g}\left[\bar{\psi}_\mu\gamma^{\mu\nu\rho}\mathcal{D}_\nu\psi_\rho
+m\bar{\psi}_\mu f^{\mu\nu}\psi_\nu\right],\qquad f^{\mu\nu}\equiv\gamma^{\mu\nu}+i(e/m^2)F^{+\mu\nu}.\eeq{s4}
The commutator of covariant derivatives is given by \beq [\mathcal{D}_\mu,\mathcal{D}_\nu]=[\nabla_\mu,\nabla_\nu]
+ieF_{\mu\nu},\eeq{s5} which, along with the relations~(\ref{g2})-(\ref{g3}), enables one to work out
the St\"uckelberg invariant Lagrangian. Thanks to the Bianchi identities and Eq.~(\ref{g5}), the result is
\bea \mathcal L_{\text{gravitino}}&=&-i\sqrt{-g}\left(\bar{\psi}_\mu\gamma^{\mu\nu\rho}\mathcal{D}_\nu\psi_\rho
+m\bar{\psi}_\mu f^{\mu\nu}\psi_\nu\right)+i\sqrt{-g}\left(\bar{\psi}_\mu f^{\mu\nu}\mathcal{D}_\nu\chi+\bar{\chi}
\overleftarrow{\mathcal D}_\mu f^{\mu\nu}\psi_\nu\right)\nonumber\\&&+\frac{e}{2m}\,\sqrt{-g}\left[F_{\mu\nu}
\left(\bar{\chi}\gamma^{\mu\nu\rho}\psi_\rho-\bar{\psi}_\rho\gamma^{\mu\nu\rho}\chi-\bar{\chi}f^{\mu\nu}\chi\right)
-\frac{1}{m}\,F_{\mu\nu}\bar{\chi}\gamma^{\mu\nu\rho}\mathcal{D}_\rho\chi\right]\nonumber\\&&-\frac{i}{2m}\,
\sqrt{-g}\left[G^{\mu\nu}\left(\bar{\chi}\gamma_\mu\psi_\nu-\bar{\psi}_\mu\gamma_\nu\chi\right)+\tfrac{1}{2}
\bar{\chi}R\chi-\frac{1}{m}\,G^{\mu\nu}\bar{\chi}\gamma_\mu\mathcal{D}_\nu\chi\right].\eea{s6}
The field redefinition that will remove the kinetic mixing is \beq \psi_\mu\rightarrow\psi_\mu
+\tfrac{1}{2}\left[\gamma_\mu-\tfrac{i}{2}(e/m^2)\,\slash{\!\!\!\!F}\gamma_\mu\right]\chi,\eeq{s7}
which can simply be found upon comparison with the model in Section 3.2. The gauge-fixing term to be added is
the appropriate version of Eq.~(\ref{em8}). Thus we are left with \bea \mathcal{L}_{\text{gravitino}}&=&-i\sqrt{-g}
\left[\,\bar{\psi}_\mu(\not{\!\!\mathcal D}-m)\psi^\mu+\tfrac{3}{2}\,\bar{\chi}(\not{\!\!\mathcal D}-m)\chi\,\right]
\nonumber\\&&+\frac{e}{2m}\,\sqrt{-g}\left[F_{\mu\nu}\left(\bar{\chi}\gamma^{\mu\nu\rho}\psi_\rho-\bar{\psi}_\rho
\gamma^{\mu\nu\rho}\chi+\bar{\chi}f^{\mu\nu}\chi\right)-\frac{1}{m}\,F_{\mu\nu}\bar{\chi}\gamma^{\mu\nu\rho}
\mathcal{D}_\rho\chi\right]\nonumber\\&&-\frac{i}{2m}\,\sqrt{-g}\left[G^{\mu\nu}\left(\bar{\chi}\gamma_\mu\psi_\nu
-\bar{\psi}_\mu\gamma_\nu\chi\right)-\tfrac{1}{2}\bar{\chi}R\chi-\frac{1}{m}\,G^{\mu\nu}\bar{\chi}\gamma_\mu
\mathcal{D}_\nu\chi\right]\nonumber\\&&+\frac{i}{2}\sqrt{-g}\,\chi\left[\left(\gamma_{\mu\nu}+\frac{ie}{m^2}
F^+_{\mu\nu}\right)\left(\gamma^\mu-\frac{ie}{2m^2}\,\slash{\!\!\!\!F}\gamma^\mu\right)+3\gamma_\nu\right]
\mathcal{D}^\nu\chi.\eea{s8} The $\mathcal{O}(F)$ contributions coming from the last line exactly cancel the
original offending operator $F_{\mu\nu}\bar{\chi}\gamma^{\mu\nu\rho}\mathcal{D}_\rho\chi$, and this can be seen
by making use of identities like~(\ref{nm13}). Then, upon the rescaling $\chi\rightarrow\sqrt{\tfrac{2}{3}}\,\chi$,
the kinetic-like operators for $\chi$ reduce to: \beq \mathcal{L}_{\chi,\text{kin}}=-i\sqrt{-g}\,\bar{\chi}
\left[g^{\mu\nu}-\frac{1}{3m^2}\left(G^{\mu\nu}+\frac{e^2}{m^2}\,F^{+\mu\rho}F_\rho^{-\,\nu}\right)\right]
\gamma_\mu\mathcal{D}_\nu\chi.\eeq{s9} Now one can use the equations of motion~(\ref{s2}) of the background
fields, and the definition~(\ref{s3}) of the EM stress-energy tensor $T^{\mu\nu}$, to rewrite the above expression
as \beq \mathcal{L}_{\chi,\text{kin}}=-i\sqrt{-g}\,\bar{\chi}\left[\left(1+\frac{\Lambda}{3m^2}\right)g^{\mu\nu}
-\frac{1}{3m^2}\left(\frac{1} {M^2_{\text P}}-\frac{2e^2}{m^2}\right)T^{\mu\nu}\right]\gamma_\mu\mathcal{D}_\nu
\chi.\eeq{s10}

If the symmetric tensor inside the brackets is proportional to the metric with a non-negative coefficient, the
$\chi$-sector will be ghost-free, and manifestly hyperbolic and causal. This is possible if the factor in front
of the stress-energy tensor is set to zero, which is nothing but imposing the charge-mass relation~(\ref{s1}).
Then, unitarity requires that the cosmological constant be bounded from below: $\Lambda\geq-3m^2$. In this
unitarily allowed region, any value of $\Lambda$ will be consistent, and in particular one can set $\Lambda=0$.

This shows that the various parameters in $\mathcal{N}=2$ (broken) supergravity~\cite{SUGRA1,SUGRA2,DZ} are tuned
precisely in a way that ensures a pathology-free helicity-1/2 sector. When $m^2=-\Lambda/3=2e^2M_{\text P}^2$, the
kinetic term~(\ref{s10}) vanishes, so that $\chi$ is relegated to a non-dynamical field. Thus we recover the
unbroken $\mathcal N=2$ AdS supergravity~\cite{SUGRA1}, where the Rarita-Schwinger field is truly massless and
enjoys null propagation.

Notice that arriving at Eq.~(\ref{s10}) from Eq.~(\ref{s9}) is a non-trivial step, and it crucially depends on
the fact that both EM and gravity are dynamical, so that the Einstein equation is sourced by the Maxwell stress-energy
tensor. This relates the two \emph{a priori} different non-canonical kinetic terms in Eq.~(\ref{s9}), and reduces their
number to one. Then the charge-mass relation removes the sole dangerous kinetic-like operator in Eq.~(\ref{s10}).
Finally, one forbids propagating ghosts in the $\chi$-sector by restricting the cosmological constant.

\section{Remarks}

The purpose of this paper was to demonstrate the power of the St\"uckelberg formalism in making transparent the
intricacies associated with interacting theories of a massive Rarita-Schwinger field. All the peculiarities $-$
such as onset of strong coupling, loss of (causal) propagation and unitarity, etc. $-$ are essentially captured in
the dynamics of the helicity-1/2 mode, and a study thereof elucidates why (in)consistent models are (in)consistent.

We have seen that EM or gravitational interactions of a massive spin-3/2 field can have a local effective field 
theory description up to energy scales parametrically larger than the mass. The finite UV cutoff signals the onset
of a dynamical regime where the helicity-1/2 sector becomes strongly coupled. Causal propagation may call for
non-minimal interactions, which could lower the intrinsic cutoff of the theory from the theoretical upper
bound reported in this paper. For example, in the case of EM coupling the required unitary-gauge Pauli term,
$i(e/m)\bar{\psi}_\mu F^{+\mu\nu}\psi_\nu$, gives rise to an $\mathcal{O}(e)$ dimension-7 operator in the
helicity-1/2 sector, and this lowers the UV cutoff to the scale: $m/\sqrt[3]{e}\ll m/\sqrt{e}$.

As pointed out in Ref.~\cite{PR1}, the cutoff scale may also mean that there could be new interacting degrees of freedom
lighter than that scale. These new fields may further improve the high energy behavior of the theory. For the
gravitational case this is exactly what happens in broken $\mathcal N=1$ supergravity~\cite{Cremmer:1978hn}.
As was shown in Ref.~\cite{sugrastrong}, a scalar and a pseudo-scalar with masses much lower than $\Lambda_\text g$
(slightly above $m$) can push the strong coupling regime all the way to the Planck scale $M_\text P$.

We have performed a St\"uckelberg analysis as a consistency check of a number of interacting theories known
in the literature. The Velo-Zwanziger acausality~\cite{VZ} of a massive spin-3/2 field minimally coupled to EM
indeed shows up as a pathology of the helicity-1/2 mode itself. ``Appropriate'' non-minimal EM interactions~\cite{PR2}
are precisely those that ensure light-cone propagation of this mode. In the case of minimal gravitational coupling,
the non-canonical kinetic terms can be rendered harmless by requiring the spacetime to be an Einstein
manifold, provided that the curvature has the well-known unitarity bound; this reconfirms the results
of~\cite{Buch,partial}. Finally, we have analyzed $\mathcal{N}=2$ (broken) supergravity~\cite{SUGRA1,SUGRA2,DZ}
to reveal that the helicity-1/2 sector acquires healthy kinetic terms in the presence of dynamical
Maxwell-Einstein fields, if the usual charge-mass relation holds.

The St\"uckelberg mode(s) can be used as a probe of the consistency of interactions for any massive particle with
$s\geq1$. While spin 2 was considered in Refs.~\cite{Nima,PR0}, it remains to be seen what more we can learn from
them about consistent interactions of massive higher spins.

\subsection*{Acknowledgments}

RR would like to thank R.~Argurio, C.~Petersson and M.~Porrati for useful discussions. He gratefully acknowledges a research 
grant from the Solvay Foundation. His work is also partially supported by IISN - Belgium (conventions 4.4511.06 and 4.4514.08),
by the Belgian Federal Science Policy Office through the Interuniversity Attraction Pole P6/11 and by the
``Communaut\'e Fran\c{c}aise de Belgique" through the ARC program, as well as by Scuola Normale Superiore, by INFN
and by the ERC Advanced Investigator Grant no.\,226455 ``Supersymmetry, Quantum Gravity and Gauge Fields" (SUPERFIELDS).

\end{document}